\documentclass[prb, twocolumn,
superscriptaddress,showpacs,amsmath,amssymb]{revtex4}
\usepackage{amsfonts}
\usepackage{bm}
\usepackage{verbatim}
\usepackage{graphicx}

\graphicspath{{pics/}}

\begin{document}

\title{Light Higgs channel of the resonant decay of magnon condensate in superfluid $^3$He-B}

\author{V.~V.~Zavjalov}
\email{vladislav.zavyalov@aalto.fi}
\affiliation{Low Temperature Laboratory, Department of Applied Physics, Aalto University, PO Box 15100, FI-00076 AALTO, Finland}

\author{S.~Autti}
\affiliation{Low Temperature Laboratory, Department of Applied Physics, Aalto University, PO Box 15100, FI-00076 AALTO, Finland}

\author{V.~B.~Eltsov}
\affiliation{Low Temperature Laboratory, Department of Applied Physics, Aalto University, PO Box 15100, FI-00076 AALTO, Finland}

\author{P.~J.~Heikkinen}
\affiliation{Low Temperature Laboratory, Department of Applied Physics, Aalto University, PO Box 15100, FI-00076 AALTO, Finland}

\author{G.~E.~Volovik}
\affiliation{Low Temperature Laboratory, Department of Applied Physics, Aalto University, PO Box 15100, FI-00076 AALTO, Finland}
\affiliation{Landau Institute for Theoretical Physics, acad. Semyonov av., 1a, 142432,
Chernogolovka, Russia}


\date{\today}

\def\omegaL{\omega_\mathrm{L}}
\def\thetaL{\theta_\mathrm{L}}
\def\OmegaB{\Omega_\mathrm{B}}
\def\Tc{T_\mathrm{c}}

\begin{abstract}
In superfluids the order parameter, which describes spontaneous symmetry breaking,
is an analogue of the Higgs field in the Standard Model of particle physics.
Oscillations of the field amplitude are massive Higgs bosons,
while oscillations of the orientation are massless Nambu-Goldstone bosons.
The 125~GeV Higgs boson, discovered at Large Hadron Collider,
is light compared to electroweak energy scale, which led to a suggestion of the
``little Higgs''  extension of the Standard Model, in which the light
Higgs appears as a NG mode acquiring mass due to violation of a hidden
symmetry. Here we show that such light Higgs exists in superfluid $^3$He-B,
where one of three Nambu-Goldstone spin-wave modes acquires small mass due to
the spin-orbit interaction. Other modes become optical and acoustic magnons.
We observe parametric decay of Bose-Einstein condensate of
optical magnons to light Higgs modes and decay of optical to acoustic
magnons. Formation of a light Higgs from a Nambu-Goldstone mode observed in
$^3$He-B opens a possibility that such scenario can be realized in
other systems, where violation of some hidden symmetry is possible,
including the Standard Model.
\end{abstract}

\maketitle

\section*{Introduction}

The superfluid transition in $^3$He, a fermionic isotope of helium, 
occurs due to formation of Cooper pairs with orbital momentum $L=1$ and
spin $S=1$. The corresponding order parameter is a $3\times3$ matrix of
complex numbers, which includes both spin and orbital degrees of
freedom.\cite{VollhardtWolfle1990} Thus, besides fermionic
quasiparticles, superfluid $^3$He possesses 18 bosonic degrees of
freedom, collective modes (oscillations) of the order parameter. Each
mode has a relativistic spectrum $\omega^2(k) = \omega_0^2 +
(c\,k)^2$, with a mode-specific wave velocity~$c$ and a gap (or
mass)~$\omega_0$. Modes with non-zero $\omega_0$ are Higgs modes
and others are Nambu-Goldstone (NG) modes. Each NG mode corresponds to
spontaneously broken continuous symmetry of the normal state.

In conventional superconductors with the order parameter of a single
complex number, only the symmetry with respect to the change of the
wave-function phase is broken. This leads to one NG phase mode and one
amplitude Higgs mode, which was experimentally
observed.\cite{Matsunaga2013,Matsunaga2014,Sherman2015} 

\begin{figure*}[t]
\centerline{\includegraphics{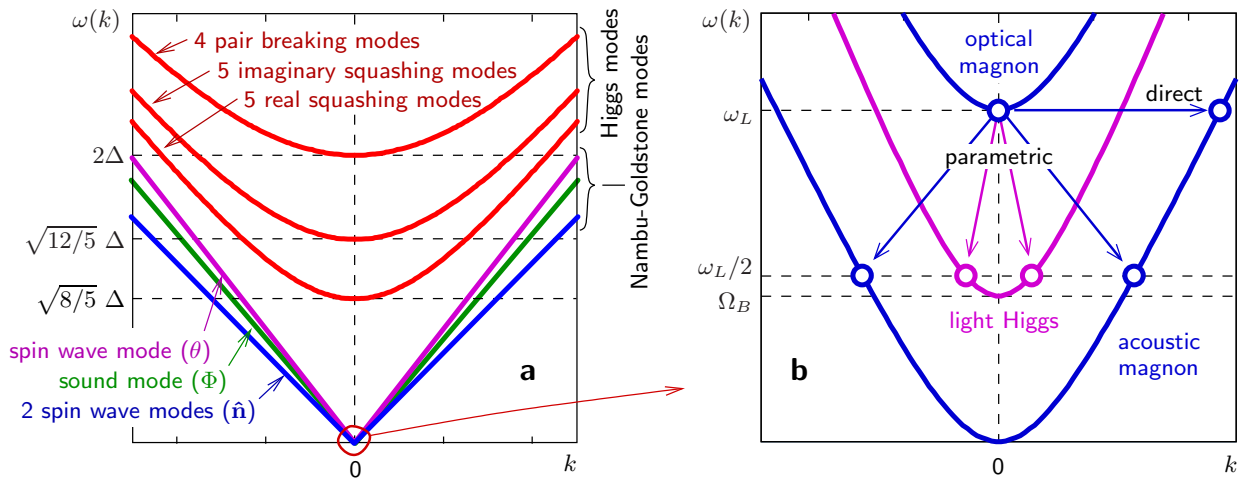}}
\caption{{\bfseries\boldmath Illustration of collective-mode spectra in $^3$He-B.}
(\textbf{a}) Modes at high energy scale ($\omega\sim\Delta\sim100\,$MHz).
From top to bottom there are six separate branches.
Three red lines show heavy Higgs modes:
four degenerate pair breaking modes with the gap~$2\,\Delta$,
five imaginary squashing modes with the gap~$\sqrt{12/5}\,\Delta$,
five real squashing modes with the gap~$\sqrt{8/5}\,\Delta$.
Nambu-Goldstone modes include a sound mode (propagating oscillations of~$\Phi$),
a spin-wave mode which corresponds to propagating oscillations of~$\theta$,
two spin-wave modes which corresponds to propagating oscillations of~$\bf\hat n$.
These modes are gapless at the high energy scale and have different propagation
velocities $c$.
(\textbf{b}) Spin-waves modes at low energy scale ($\omega\sim10^{-3}\Delta\sim100\,$kHz).
Due to spin-orbit interaction the $\theta$ mode acquires a small gap $\OmegaB \ll \Delta$
and becomes a light Higgs boson.
Two $\hat{\mathbf{n}}$ modes are split by the magnetic field into optical
and acoustic magnons. Arrows indicate decay channels observed in our experiments.}
\label{image:he3modes}
\end{figure*}

In unconventional B phase of superfluid $^3$He the symmetry with respect
to relative rotations of the spin and orbital spaces is additionally
broken, \cite{Leggett1975} and the order parameter in the zero magnetic
field is
\begin{equation} \label{Aalphai}
A_{\alpha i}=
\Delta ~e^{i\Phi}~R_{\alpha i}({\bf\hat n},\theta)\,,
\end{equation}
where $\Delta$ is the gap in the fermionic spectrum,~$\Phi$ is the phase
and~$R_{\alpha i}$ is a rotation matrix, which connects spin and orbital
degrees of freedom. The matrix~$R_{\alpha i}$ is represented in terms of
the rotation axis~$\bf\hat n$ and  angle~$\theta$. Parameters $\Phi$,
${\bf\hat n}$ and $\theta$ determine a 4-dimensional  subspace of
degenerate states. Thus among 18 collective modes of $^3$He-B,
Fig.~\ref{image:he3modes}a, four are NG modes: oscillation of~$\Phi$ is
sound and oscillations of~$R_{\alpha i}$ (or~$\bf\hat n$ and~$\theta$)
are spin waves. The other 14 modes are the Higgs modes with energy gaps
of the order of $\Delta$. These heavy Higgs modes have been investigated
for a long time both theoretically\cite{Vdovin1963, Maki1974, Nagai1975,
Tewordt-Einzel1976} and experimentally.\cite{Giannetta1980, Mast1980,
Avenel1980, Lee1988, Collett2012} 


In superconductors and in the Standard Model, the NG bosons become
massive due to the Anderson-Higgs mechanism.\cite{Anderson1963,
Higgs1964, PekkerVarma2014} In electrically neutral $^3$He  these modes
are gapless, when viewed from the scale of $\Delta \sim 100\,$MHz (set
by the critical temperature $\Tc \sim 10^{-3}\,$K). At low energy
scale of $\sim1\,$MHz, corresponding to the frequency of our nuclear
magnetic resonance (NMR) experiments, two weak effects become
significant: spin-orbit interaction and applied magnetic field~$\bf H$,
Fig.~\ref{image:he3modes}b.

Spin-orbit interaction lifts the degeneracy with respect to $\theta$,
and the minimum energy corresponds to the so-called Leggett angle
$\thetaL = \arccos(-1/4)$. This explicit violation of the  symmetry of
the B phase leads to appearance of the gap for the $\theta$ mode. This
mode becomes an additional, light, Higgs boson. The  gap value~$\OmegaB$
is called Leggett frequency, it is a measure of the spin-orbit
interaction. At low temperatures $\OmegaB\sim 100\,$kHz $\ll \Delta$.

Two other spin-wave modes are oscillations of ${\bf\hat n}$. In the
magnetic field the equilibrium state corresponds to ${\bf\hat
n}\parallel {\bf H}$ and the field splits these two modes  in the same
way as in ferromagnets. One of the modes, the optical magnon, acquires
the gap equal to the Larmor frequency $\omegaL = \gamma H$
(where~$\gamma$ is the  gyromagnetic ratio). Another one, the acoustic
magnon, remains gapless, but its spectrum becomes quadratic. Such
unusual form of the spectrum comes from a violation of the time reversal
symmetry by the magnetic field (general discussion of the NG modes with
quadratic spectrum  see in Ref.~\onlinecite{Nitta2015}).

All three low frequency spin-wave modes are described by the closed
system of Leggett equations.\cite{VollhardtWolfle1990}  In particle
physics such set of low-energy modes, which includes NG modes and a
light Higgs, is called the Little Higgs
field.\cite{LittleHiggsReview2005} Formal definition of such field in
$^3$He-B is given in Supplementary Note 1. 

In an NMR experiment one follows the dynamics of magnetization
$\mathbf{M}$, or of the spin $\mathbf{S}$ of the sample. The motion
of~$\theta$, corresponds to longitudinal spin waves, $\delta {\bf S}
\parallel {\bf H}$, while oscillations of~$\bf\hat n$ correspond to
transverse spin motion, $\delta {\bf S} \perp {\bf H}$. Optical magnons
can be directly created with traditional transverse NMR. With a suitable
coil system one can also directly excite longitudinal spin oscillations,
or light Higgs mode.\cite{OsheroffNMR} Coupling to short-wavelength
acoustic magnons is hard to achieve in a traditional NMR experiment with
large excitation coils. 

In this work we use a technique, based on Bose-Einstein condensate (BEC)
of optical magnons,\cite{BunkovVolovik2013} to probe interaction and
conversion between all components of the little Higgs field in $^3$He-B.
As a result, we observe parametric decay of optical magnons to light
Higgs bosons, and both parametric and direct conversion between optical
and acoustic magnons. The measured mass of light Higgs and propagation
velocity of acoustic magnons are close to the expected values. Thus we
experimentally confirm the little Higgs scenario in $^3$He-B. The little
Higgs field  appears in quantum chromodynamics,\cite{Dobrescu2014} where
NG modes (pions) acquire light mass due to the explicit violation of the
chiral symmetry, which is negligible at high energy, but becomes
significant at low energy.\cite{Weinberg1972} The relatively small mass
of the 125~GeV Higgs boson observed at the Large Hadron Collider suggests
that it might be also the pseudo-Goldstone (light Higgs) boson (see e.g.
Ref.~\onlinecite{VolovikZubkov2014b} and references therein).


\begin{figure}[b]
\centerline{\includegraphics{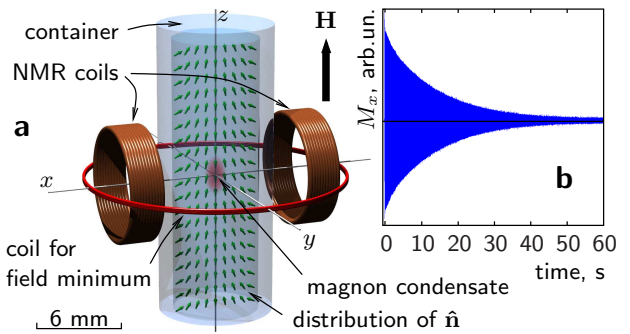}}
\caption{\textbf{Experimental setup.}
(\textbf{a})  Superfluid $^3$He-B is confined in a cylindrical quartz
container, and a constant magnetic field $\mathbf{H}$ is applied along
the container axis. A special coil creates a minimum of the field
magnitude $H$ in the axial direction, while transverse NMR coils are used
to pump optical magnons and to detect magnetization precession. Green
arrows show equilibrium distribution of the ${\bf\hat n}$ vector, which
together with the $\mathbf{H}$ profile creates a trap for optical magnons
near the axis of the sample. In this trap magnon BEC is formed.
(\textbf{b}) An example of the signal from the NMR coil during
the condensate decay measured at $\omega/2\pi=833$~kHz and $P=0$~bar.
Its amplitude is proportional to the coherently precessing transverse
magnetization of the condensate.}
\label{image:cell}
\end{figure}

\begin{figure*}[t]
\centerline{\includegraphics{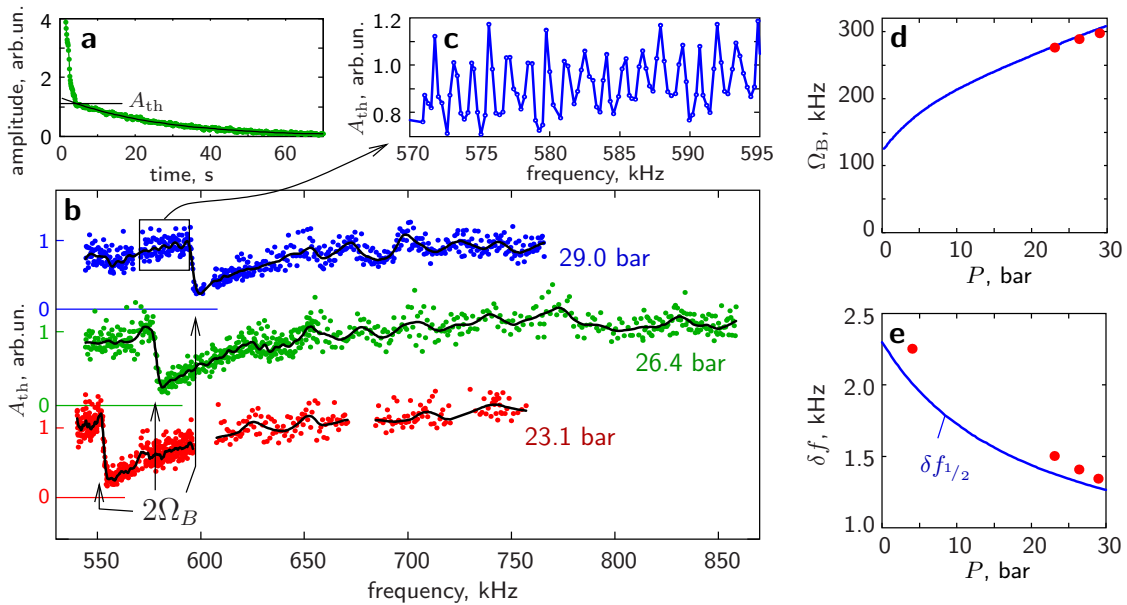}}
\caption{\textbf{Measurements of the Suhl instability.}
(\textbf{a}) An example measurement of the NMR signal amplitude during
the magnon condensate decay obtained at $\omega_{\rm L}/2\pi = 741$~kHz and
$P=26.4$~bar. A Suhl instability threshold at the amplitude $A_{\rm
  th}$ is clearly seen.
(\textbf{b}) Dependence of the threshold amplitude~$A_{\rm th}$ on the frequency of
optical magnons $\omega_{\rm opt} \approx \omegaL$ (colored symbols) for
three pressures $P$. Sharp decrease is seen at $\omega_{\rm opt} =
2\OmegaB(P)$, where the decay of an optical magnon to two light Higgs
bosons becomes possible. Black lines are smoothed data using running
average. Curves for different pressures are shifted in the vertical
direction and respective zero levels are marked by horizontal lines.
(\textbf{c}) Zoom to one measurement in the panel~\textbf{b}
shows periodic modulation, which corresponds to resonances
of acoustic magnons in the experimental container.
(\textbf{d}) Pressure dependence of the light Higgs boson mass $\OmegaB$
(symbols) from measurements in the panel~\textbf{b}.
(\textbf{e}) Frequency separation of acoustic magnon resonances
(symbols) from measurements like in the panel~\textbf{c}. Solid lines in
the panels~\textbf{d} and~\textbf{e} are theoretical values based on known
$^3$He parameters\cite{2001_thuneberg} without fitting. }
\label{image:suhl}
\end{figure*}

\section*{Results}

\textbf{Suhl instability.}
As a tool to study dynamics of the little Higgs field in superfluid
$^3$He-B we use trapped Bose-Einstein condensates of optical magnons,
Fig.~\ref{image:cell}. The condensate is well separated from the
container walls, where the strongest magnetic relaxation in $^3$He
usually occurs.\cite{Lanc-decay} Thus tiny relaxation effects, connected
to coupling of optical magnons to other components of the little Higgs
field, can be observed.

When number of pumped magnons is low, slow exponential relaxation of the
precession signal is determined by spin diffusion and energy losses in
the NMR pick-up circuit,\cite{MagnonRelaxation} Fig.~\ref{image:cell}b.
We have found that above some threshold amplitude the relaxation becomes much
faster, Fig.~\ref{image:suhl}a. The explanation is  the Suhl
instability,\cite{Suhl1957} a well-known nonlinear effect in magnets when
a uniform precession of magnetization (here at the optical magnon
frequency $\omega_{\rm opt}$) parametrically excites a pair of acoustic
magnons with twice smaller frequency $\omega_{\rm ac}$ and opposite ${\bf
k}$-vectors: $\omega_{\rm opt} = \omega_{\rm ac}({\bf k})+\omega_{\rm
ac}(-{\bf k})$. In the case of $^3$He-B both acoustic magnons and the
light Higgs modes can be parametrically  excited,
Fig.~\ref{image:he3modes}b. The process  occurs with conservation of
energy and momentum. The threshold amplitude is inversely proportional to
the coupling between  decaying and excited waves and proportional to the
relaxation in the excited wave.

\textbf{Mass of light Higgs.}
The measured threshold amplitude as a function of NMR frequency and
pressure is plotted in Fig.~\ref{image:suhl}b.  The frequency dependence
allows us to identify the decay channels. It is clear from
Fig.~\ref{image:he3modes}b that the decay of the optical magnon to a pair
of light Higgs bosons with the frequency $\omega_{\rm Higgs}$,
$\omega_{\rm opt} = \omega_{\rm Higgs}({\bf k})+\omega_{\rm Higgs}(-{\bf
k})$, is possible only when the precession frequency is larger then
$2\OmegaB$. We see a pronounced drop of the threshold amplitude at this
frequency: The threshold decreases by about an order of magnitude.  

In Fig.~\ref{image:suhl}d the measured mass of light Higgs $\OmegaB$ is
plotted as a function of pressure. Measurements are in a good agreement
with values of the Leggett frequency from
Ref.~\onlinecite{2001_thuneberg}.

\textbf{Resonances of acoustic magnons.}
In addition to the sharp drop, connected with light Higgs mode, we find
periodic modulation of the threshold amplitude as a function of the
frequency of the precession, Fig.~\ref{image:suhl}c. These periodic peaks
originate from the parametric decay of the optical magnons in the BEC to
acoustic magnons. The frequency dependence is explained by quantization
of the magnon spectrum in the cylindrical container, which serves as a
resonator for acoustic magnons. Consider a decay of the optical magnon
with frequency~$\omega_{\rm opt}$ into acoustic magnons with
frequency~$\omega_{\rm ac}=N\,\omega_{\rm opt}$, where for the parametric
excitation $N=1/2$. In the following discussion we will use $\omega_{\rm
opt}=\omegaL$ since the difference is negligible for trapped optical
magnons. By sweeping the magnetic field we can change both magnon spectrum
and magnon trap and observe resonances in the cell. The simple
resonance condition for acoustic magnons in a cylinder with the
radius~$R$ gives the distance between the resonances (Supplementary Note 2):
\begin{equation}\label{eq:simple_periods}
\delta f_N = \frac{1}{\sqrt{N(1+N)}}\ \frac{c}{4R}\,.
\end{equation}
where $c$ is the relevant spin-wave velocity.

In Fig.~\ref{image:suhl}e the measured acoustic magnon resonance
period~$\delta f_{^1\!/{}_2}$ is plotted as a function of pressure. The
results are in a good agreement with Eq.~(\ref{eq:simple_periods}), where
values of the spin-wave velocity are taken from our recent
measurements.\cite{SpinWaveVel}

\begin{figure}[b]
\centerline{\includegraphics{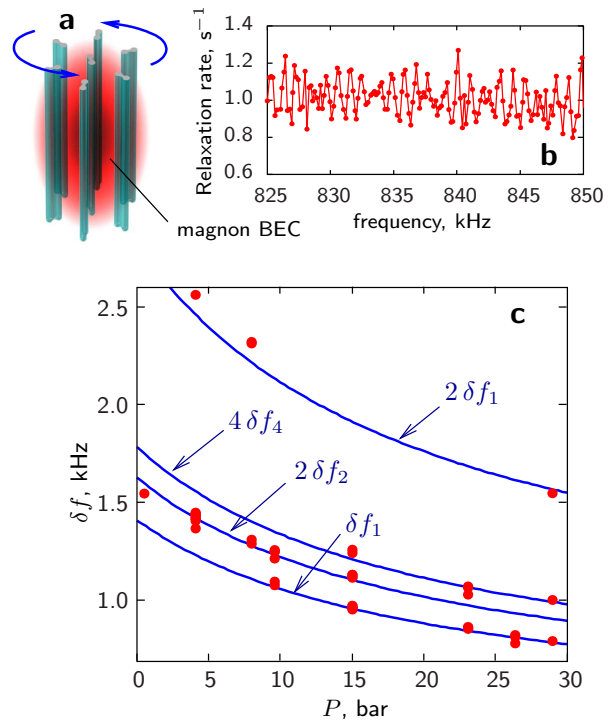}}
\caption{\textbf{Excitation of acoustic magnons by vortices.}
(\textbf{a}) Schematic plot of vortices in rotating $^3$He-B. Precession
of magnetization of the magnon condensate cause
oscillations of non-axisymmetric vortex cores. The oscillations produce
acoustic magnons and increase relaxation of the condensate.
(\textbf{b}) Relaxation rate of the magnon condensate as a function of
frequency, measured at $\Omega=1$~radian per second and $P=23.4$~bar. Two sets of
acoustic magnon resonances with periods about~$1$~kHz produce a clearly
seen beat with a period of~$5$~kHz. 
(\textbf{c}) Measured periods of magnon condensate relaxation peaks
(symbols) as a function of pressure. Lines are plotted using
equation~(\ref{eq:simple_periods}) for acoustic magnon resonances without
fitting parameters. }
\label{image:vort}
\end{figure}

\textbf{Effect of quantized vortices.}
An additional relaxation mechanism for the magnon condensate is found
when quantized vortices are formed in the sample. In the presence of
these localized topological objects the momentum ${\bf k}$ of
the spin-wave modes is not conserved,  and one expects direct
excitation of acoustic magnons by the optical mode. We can rotate the
sample with angular velocities up to $\Omega=2$ radian per second to create a
cluster of rectilinear quantized  vortices, which cross the
whole experimental region including the magnon BEC,
Fig.~\ref{image:vort}a.  In this state the
relaxation rate, plotted as a function of the frequency in
Fig.~\ref{image:vort}b, reveals several periodic sets of peaks. We
attribute these peaks to resonances of acoustic magnons with
frequencies
$\omegaL$, $2\omegaL$, etc.

In $^3$He-B the rotational symmetry of a vortex is spontaneously broken
and the vortex core can be treated as a bound state of two half-quantum
vortices which can rotate around the vortex axis. Dynamics of the vortex
is affected by the precessing magnetization
\cite{VortexTrans,SilaevThuneberg}. Precession of ${\bf  S}$ and
${\bf\hat n}$ in the magnon BEC produces torsional oscillations of the
vortex core. The fact that the equilibrium position of~$\bf\hat n$
deviates from the vertical direction within the magnon BEC makes this
oscillations unharmonic. As a result, acoustic magnons with frequencies
$N\omegaL$ can be emitted.

The amplitudes of the various resonances depend on a distribution
of vortex cores and the wave nodes of acoustic magnons. For example, an
axially symmetric distribution of vortices can excite only symmetric
waves, which means doubling of the observed resonance period. In our
experiment acoustic magnons (with wave length~$5-10$~$\mu$m) are emitted
by vortices, which are within the magnon condensate.  The distance
between vortices ~$0.1-0.2$~mm is comparable with the size of the
trapped condensate $0.2-0.4$~mm. Thus the amplitudes of resonances are
sensitive to details, such as order parameter texture, rotation and
pressure and we do not see all the harmonics at all pressures.
Nevertheless the resonance periods plotted in
Fig.~\ref{image:vort}c follow the theoretical
values~(\ref{eq:simple_periods}) or their multiples (denoted as
$2\,\delta f_1$, etc.).


\section*{Discussion}

To summarize, we have observed the interplay of all three spin wave
modes, which form a little Higgs field in superfluid~$^3$He-B. In
particular, we have found two channels of parametric decay of optical
magnons: to a pair of light Higgs bosons and to a pair of acoustic
magnons.  While the search for similar resonant production of pairs of
Standard Model Higgs bosons reported by the ATLAS
collaboration~\cite{ATLAS2015} has not succeeded yet, our results
support the basic physical idea behind this effort. Another system where
the light Higgs mode can be observed is the multicomponent condensate in
cold gases~\cite{ueda}, where interaction between components can be set
up to produce the hidden symmetry.

We find that the low-energy physics in superfluid $^3$He has many common
features of the Higgs scenario in Standard Model: both are described by
the SU(2) and U(1) symmetry groups; the acoustic and optical magnons
correspond to the  doublet of W$^+$ and W$^-$  gauge bosons which
spectrum also splits in magnetic field;\cite{Chernodub2014} the light
Higgs mode has parallel with the 125\,GeV Higgs boson. However, in
addition, the $^3$He-B has the high-energy sector with 14 heavy Higgs
modes. This suggests that in the same manner the 125\,GeV Higgs boson
belongs to the low energy sector of particle physics, and if so, one may
expect the existence of the heavy Higgs bosons at TeV scale.

We have demonstrated that the short-wavelength acoustic magnons can be
emitted and detected with the BEC of optical magnons. Acoustic magnons
can be lensed by non-uniform magnetic fields and the order-parameter
texture, and thus might serve in future as a powerful local probe to
study topological superfluidity of $^3$He, including Majorana fermions on
the boundaries of the superfluid and in the cores of quantized vortices.

\section*{Methods}

\small
\textbf{The sample geometry and NMR setup.} Superfluid $^3$He-B
is placed in a cylindrical container with the inner diameter of
$2R=5.85$\,mm, made from fused quartz, Fig.~\ref{image:cell}. The
container has closed top end and open bottom end, which provides the
thermal contact to the nuclear demagnetization refrigerator. Static
magnetic field is applied parallel to the container axis. A special coil
creates a controlled minimum of the field magnitude along the axial
direction. Transverse NMR coils, made from copper wire, are used to
create and detect magnetization precession. Coils are part of a tuned
tank circuit with the Q value of around 130. Frequency tuning is provided
by a switchable capacitance bank, installed at the mixing chamber of the
dilution refrigerator. To improve signal to noise ratio, we use a cold
preamplifier, thermalized to liquid helium bath. 

The measurements are performed at low temperatures $T < 0.2\ \Tc$, where
spin-wave velocities and the Leggett frequency are
temperature-independent. Typically we use $T = 130 - 350\,\mu$K,
depending on pressure. The temperature is measured by a quartz tuning
fork thermometer, installed at the bottom of the sample cylinder. The
heat leak to the sample was measured in earlier work to be about
$12\,$pW.\cite{thermalPRB} The measurements are performed at pressures
$0-29$~bar and in magnetic fields $H=17-26$~mT with corresponding NMR
frequencies $\omegaL/2\pi=550-830$~kHz. 

\textbf{Magnon trap.}
Minimum of the axial magnetic field forms a trapping potential for
optical magnon quasiparticles in the axial direction. Trapping in the
radial direction is provided by the spin-orbit interaction via the
equilibrium distribution of the order parameter. In this geometry it
forms the so-called flare-out texture: $\bf\hat n$ is parallel to $\bf H$
on the cell axis and tilted near walls because of boundary
conditions.\cite{Smith1977} The combined magneto-textural trap is nearly
harmonic with trapping length about 0.3\,mm in the radial direction and
1\,mm in the axial direction (see Supplementary Note 3 for details).

\textbf{Measurements of magnon BEC.}
Owing to the geometry, the coils couple only to optical magnons with
$k\approx0$.  With a short rf pulse in the NMR coils non-equilibrium
optical magnons are created. At temperatures of our experiment
equilibration within the magnon subsystem proceeds much faster that the
decay of magnon number, and the pumped magnons are condensed to the
ground level of the trap within 0.1\,s from the pulse. Manifestation of
Bose-Einstein condensation is the spontaneously coherent precession of
the condensate magnetization,\cite{lancOffRes,ourOffRes} which induces
current in the NMR coils. The amplified signal is recorded by a digital
oscilloscope; an example record is in Fig.~\ref{image:cell}b. We then
perform sliding Fourier transform of the signal with the window
$0.3-1\,$s. In the resulting sharp peak in the spectrum the frequency
determines the BEC precession frequency $\omega_{\rm opt}$, while the
amplitude (such as shown in Fig.~\ref{image:suhl}a) is proportional to
the square root of the number of magnons in the trap.

\textbf{Rotation.}
The sample is installed in the rotating nuclear demagnetization
refrigerator ROTA,\cite{ROTA} and can be put in rotation together with
the cryostat and the measuring equipment. The cryostat is properly
balanced and suspended on active vibration isolation, and in rotation
the heat leak to the sample remains below 20\,pW.\cite{thermalPRB}
Vortices are created by increasing angular velocity $\Omega$ from zero
to a target value at temperature around $0.7\,\Tc$, where the
mutual friction allows for fast relaxation of vortex configuration
towards an equilibrium array.\cite{PNAS} Further cool-down is performed
in rotation.

\bibliographystyle{naturemag}

\begin{thebibliography}{15}

\bibitem{VollhardtWolfle1990}
Vollhardt, D. \&  W\"olfle, P.
{\it The superfluid phases of helium 3},
Taylor and Francis, London  (1990).

\bibitem{Matsunaga2013}
Matsunaga, R. \textit{et al.} 
The Higgs amplitude mode in BCS superconductors
Nb$_{1-x}$Ti$_{x}$N induced by terahertz pulse excitation.
\textit{Phys. Rev. Lett.} {\bf 111}, 057002 (2013).

\bibitem{Matsunaga2014}
Matsunaga, R. \textit{et al.}
Light-induced collective pseudospin precession resonating with Higgs mode in a superconductor.
\textit{Science} {\bf 345}, 1145--1149  (2014).

\bibitem{Sherman2015}
Sherman, D. {\it et al.}
The Higgs mode in disordered superconductors close to a quantum phase transition.
\textit{Nature Phys.} {\bf 11}, 188--192 (2015).

\bibitem{Leggett1975}
Leggett, A.J.
A theoretical description of the new phases of liquid $^3$He,
\textit{Rev. Mod. Phys.} {\bf 47}, 331--414 (1975).

\bibitem{Vdovin1963}
Vdovin, Y.A.
in: {\it Applications of Methods of Quantum Field Theory to Many Body
Problems}, ed. A.I. Alekseyeva (GOS ATOM IZDAT, Moscow, 1963).

\bibitem{Maki1974}
Maki, K.
Propagation of zero sound in the Balian-Werthamer state.
\textit{J. Low Temp. Phys.} \textbf{16}, 465--477 (1974).

\bibitem{Nagai1975}
Nagai, K.
Collective excitations from the Balian-Werthamer state.
\textit{Prog. Theor. Phys.}  \textbf{54},  1--18 (1975).

\bibitem{Tewordt-Einzel1976}
Tewordt, L. \& Einzel, D.
Collective modes and gap equations for the superfluid states in $^3$He.
\textit{Phys. Lett. A} \textbf{56},   97--98 (1976).


\bibitem{Giannetta1980}
Giannetta, R. et al.,
Observation of a New Sound Attenuation Peak in Superfluid $^3$He-B.
\textit{Phys. Rev. Lett.} \textbf{45}, 262--265 (1980);

\bibitem{Mast1980}
Mast, D. et al.,
Measurements of High-Frequency Sound Propagation in $^3$He-B.
\textit{Phys. Rev. Lett.} \textbf{45}, 266--269 (1980).

\bibitem{Avenel1980}
Avenel, O., Varoquaux, E. \& Ebisawa, H.
Field splitting of the new sound attenuation peak in $^3$He-B.
\textit{Phys. Rev. Lett.} \textbf{45}, 1952--1955 (1980).

\bibitem{Lee1988}
Movshovich, R., Varoquaux, E., Kim, N. \& Lee, D.M.
Splitting of the squashing collective mode of superfluid
$^3$He-B by a magnetic field.
\textit{Phys. Rev. Lett.} \textbf{61}, 1732--1735 (1988).

\bibitem{Collett2012}
Collett, C. A., Pollanen, J., Li, J. I. A., Gannon, W. J. \& Halperin, W. P.
Zeeman splitting and nonlinear field-dependence in superfluid $^3$He.
\textit{J. Low Temp. Phys.} \textbf{171},  214--219 (2013).



\bibitem{Higgs1964}
Higgs, P.W. 
Broken Symmetries and the Masses of Gauge Bosons.
\textit{Phys. Rev. Lett.} {\bf 13}, 508--509 (1964).

\bibitem{Anderson1963}
Anderson, P. W.
Plasmons, gauge invariance, and mass.
\textit{Phys. Rev.} {\bf 130}, 439--442 (1963).

\bibitem{PekkerVarma2014}
Pekker D. \&  Varma, C.M.
Amplitude/Higgs modes in condensed matter physics.
\textit{Annu. Rev. Condens. Matter Phys.} {\bf 6} 269--297 (2015).



\bibitem{Nitta2015}
Nitta, M. \& Takahashi, D.A.
Quasi-Nambu-Goldstone modes in nonrelativistic systems.
\textit{Phys. Rev. D} {\bf 91}, 025018 (2015).
%

\bibitem{LittleHiggsReview2005}
Schmaltz, M. \& Tucker-Smith, D.
Little Higgs Review.
\textit{Ann. Rev. Nucl. Part. Sci.} {\bf 55}, 229--270 (2005).

\bibitem{OsheroffNMR}
Osheroff, D. D.
Longitudinal and Transverse Resonance in the B Phase of Superfluid $^3$He.
\textit{Phys. Rev. Lett.} {\bf 33}, 1009--1012 (1974).

\bibitem{BunkovVolovik2013}
Bunkov, Yu. M. \& Volovik, G. E.
Spin superfluidity and magnon BEC.
\textit{International Series of Monographs on Physics} {\bf 156}, Volume 1, 253--311 (2013).

\bibitem{Dobrescu2014}
Dobrescu, B.A. \&  Frugiuele, C.
Hidden GeV-scale interactions of quarks.
\textit{Phys. Rev. Lett.} {\bf 113}, 061801 (2014).

\bibitem{Weinberg1972}
Weinberg, S.
Approximate symmetries and pseudo-Goldstone bosons.
\textit{Phys. Rev. Lett.} {\bf 29}, 1698--1701 (1972).

\bibitem{VolovikZubkov2014b}
Volovik, G.E. \& Zubkov, M.A.
Scalar excitation with Leggett frequency in $^3$He-B and the $125$ GeV Higgs particle in top quark condensation models as pseudo-Goldstone bosons.
\textit{ Phys. Rev.} D {\bf 92}, 055004 (2015).



%

\bibitem{Lanc-decay}
Fisher, S. N., Pickett, G. R., Skyba, P. \& Suramlishvili, N.
Decay of persistent precessing domains in $^3$He-B at very low temperatures.
Phys. Rev. B \textbf{86}, 024506 (2012).

\bibitem{MagnonRelaxation}
Heikkinen, P.J., Autti, S., Eltsov, V.B., Haley, R.P. \& Zavjalov, V.V.
Microkelvin thermometry with Bose-Einstein condensates of magnons and
applications to studies of the AB interface in superfluid $^3$He.
\textit{J. Low Temp. Phys.} {\bf 175}, 681--705 (2014).

%

\bibitem{Suhl1957}
Suhl, H.
The theory of ferromagnetic resonance at high signal powers.
\textit{J. Phys. Chem. Solids} {\bf 1}, 209--227 (1957).

\bibitem{2001_thuneberg}
Thuneberg, E.V.
Hydrostatic theory of superfluid $^3$He-B.
\textit{J. Low Temp. Phys.} {\bf 122}, 657--682 (2001).

%

\bibitem{SpinWaveVel}
Zavjalov, V.V., Autti, S., Eltsov, V.B. \& Heikkinen, P.J.
Measurements of the anisotropic mass of magnons confined in a harmonic trap in superfluid $^3$He-B
\textit{JETP Letters} {\bf 101}, 802-807 (2015)

\bibitem{VortexTrans}
Kondo, Y. \textit{et al.}
Direct observation of the nonaxisymmetric vortex in superfluid $^3$He-B.
\textit{Phys. Rev. Lett.} {\bf 67}, 81--84 (1991).

\bibitem{SilaevThuneberg}
Silaev, M.A., Thuneberg, E.V. \& Fogelstr\"om, M.
Lifshitz transition in the double-core vortex in $^3$He-B.
arXiv:1505.02136 (2015).

\bibitem{ATLAS2015}
Aad, G. \textit{et al.} (ATLAS Collaboration)
Search for Higgs boson pair production in the $\gamma\gamma bb^-$ final state using pp collision data at $\sqrt{s}=8$ TeV from the ATLAS detector.
\textit{Phys. Rev. Lett.} {\bf 114}, 081802 (2015).

\bibitem{ueda}
Ueda M.,
{\it Fundamentals and New Frontiers of Bose-Einstein Condensation.}
World Scientific, 2010.

\bibitem{Chernodub2014}
Chernodub, M.N.
Superconducting properties of vacuum in strong magnetic field,
\textit{Int. J. Mod. Phys. D} {\bf 23}, 1430009 (2014).


\bibitem{thermalPRB}
Hosio J.J. \textit{et al.}
Propagation of thermal excitations in a cluster of vortices in superfluid $^3$He-B.
\textit{Phys. Rev. B} \textbf{84}, 224501 (2011).

\bibitem{Smith1977}
Smith, H., Brinkman, W. F. \& Engelsberg, S.
Textures and NMR in superfluid $^3$He-B.
\textit{Phys. Rev. B} {\bf 15}, 199--213 (1977).

\bibitem{lancOffRes}
Cousins, D.J., Fisher, S.N., Gregory, A.I., Pickett, G.R. \& Shaw, N.S.
Persistent coherent spin precession in superfluid $^3$He-B driven by off-resonant excitation.
\textit{Phys. Rev. Lett.} {\bf 82}, 4484--4487 (1999).

\bibitem{ourOffRes}
Autti, S. \textit{et al.}
Self-trapping of magnon Bose-Einstein condensates in the ground state and   on excited levels: From harmonic to box confinement.
\textit{Phys. Rev. Lett.} {\bf 108}, 145303 (2012).

\bibitem{ROTA}
Hakonen, P.J. et al.,
Rotating nuclear demagnetization refrigerator for experiments on superfluid He$^3$.
\textit{Cryogenics} {\bf 23}, 243--250 (1983).

\bibitem{PNAS}
Eltsov, V.B., H\"anninen, R. \& Krusius, M.
Quantum turbulence in superfluids with wall-clamped normal component.
\textit{Proc. Natl. Acad. Sci. USA} \textbf{111}, 4711--4718 (2014).

%
%
%
%
%
%

%
%
%
%
%

\end{thebibliography}

\begin{thebibliography}{10}
\section*{SUPPLEMENTARY REFERENCES}
\bibitem{2001_thuneberg}
Thuneberg, E.V.
Hydrostatic theory of superfluid $^3$He-B.
\textit{J. Low Temp. Phys.} {\bf 122}, 657--682 (2001).

\bibitem{1983_theodorakis}
Theodorakis, S. \& Fetter, A.L.
Vortices and NMR in rotating $^3$He-B.
\textit{J. Low Temp. Phys.} {\bf 52}, 559--591 (1983).

\bibitem{Chernodub2014}
Chernodub, M.N.
Superconducting properties of vacuum in strong magnetic field,
\textit{Int. J. Mod. Phys. D} {\bf 23}, 1430009 (2014).

\bibitem{SpinWaveVel}
Zavjalov, V.V., Autti, S., Eltsov, V.B. \& Heikkinen, P.J.
Measurements of the anisotropic mass of magnons confined in a harmonic trap in superfluid $^3$He-B.
\textit{JETP Letters} {\bf 101}, 802--807 (2015).

\end{thebibliography}

\section*{Acknowledgements}
We thank M. Krusius and V.S L'vov for useful discussions.
This work has been supported in part by the EU 7th Framework Programme 
(FP7/2007-2013, Grant No. 228464 Microkelvin), by the Academy of Finland
(project no. 284594), and by the facilities of the Cryohall
infrastructure of Aalto University. P.J.H. acknowledges financial support
from the V\"{a}is\"{a}l\"{a} Foundation of the Finnish Academy of Science
and Letters, and S.A. that from the Finnish Cultural Foundation.

\section*{Author contributions}

The experiments were conducted by S.A., P.J.H., V.V.Z., and V.B.E., the
theoretical analysis was carried out by G.E.V and V.V.Z., the paper was
written by V.V.Z., G.E.V. and V.B.E., with contribution from all the
authors.

\section*{Conflicting financial interests}
The authors declare no competing financial interests.

\newpage
\section*{SUPPLEMENTARY NOTE 1}

\subsection*{Spin waves in $^3$He-B}
\def\ts{\theta^s}
\def\ct{\cos\theta}
\def\st{\sin\theta}
\def\ddd#1#2{\frac{\partial #1}{\partial x_{#2}}}
\def\FSO{F_\mathrm{so}}

Spin waves in$^3$He-B correspond to motions of the rotation matrix
$R_{aj}$. The matrix can be represented by means of the rotation
axis~$\bf\hat n$ and the rotation angle~$\theta$ as
\begin{equation}\label{eq:r_nt}
R_{a j} = \ct\ \delta_{a j} + (1-\ct)\ n_a n_j - \st\ e_{ajk} n_k.
\end{equation}

The motion is affected by the energy of the spin-orbit interaction $\FSO$
and the gradient energy $F_\nabla$:
\begin{eqnarray}
\label{eq:en_d}
\FSO &=& \frac{\chi_B\OmegaB^2}{15\gamma^2}  (R_{jj}R_{kk} + R_{jk}R_{kj}),\\
\label{eq:en_g}
F_\nabla
&=& \frac12 \Delta^2 ( K_1 G_1 + K_2 G_2 + K_3 G_3),
\end{eqnarray}
where
\begin{eqnarray*}
G_1 &=& \nabla_j R_{ak} \nabla_j R_{ak},\\
G_2 &=& \nabla_j R_{ak} \nabla_k R_{aj},\\
G_3 &=& \nabla_j R_{aj} \nabla_k R_{ak},
\end{eqnarray*}
$\chi_B$ is the spin susceptibility of the $^3$He-B, ${\gamma}$ the
gyromagnetic ratio for the $^3$He atom, $\OmegaB$ the Leggett frequency,
$\Delta$ the superfluid gap, and $K_1, K_2$ and $K_3$ are parameters of
the gradient energy.

The spin-orbit interaction energy has a simple form in terms of~$\bf\hat n$
and~$\theta$ with a minimum at~$\theta=\arccos(-1/4)$:

\begin{equation}\label{eq:en_d_nt}
\FSO = \frac{\chi_B\OmegaB^2}{15\gamma^2} \left(\ct+1/4\right)^2 + \mathrm{const.}
\end{equation}

The equation of small spin oscillations near the equilibrium value
${\bf S}^0=(\chi_B/\gamma)\ {\bf H}$ is~\cite{1983_theodorakis}
\begin{eqnarray}\label{eq:ham_eq3}
\ddot S_c &=& [\dot {\bf S}\times \gamma {\bf H}]_c \\\nonumber
&+& \frac{\Delta^2\gamma^2}{\chi_B} \left[
K\ \nabla^2\ S_c
- K'\ \nabla_j R^0_{cj} R^0_{ak} \nabla_k\ S_a\right]\\\nonumber
&-& \OmegaB^2\ {\bf\hat n}\cdot ({\bf S} - {\bf S^0})\ \hat n_c,
\end{eqnarray}
where~$K=2K_1+K_2+K_3$ and~$K'=K_2+K_3$.

In a texture with ${\bf\hat n} \parallel {\bf H}$ or in a high magnetic
field $\omegaL^2/\OmegaB^2 \gg 1$ one can separate transverse and
longitudinal oscillations. In the case of short wavelengths (when the
spin changes on a much shorter distance than the texture) one can write
the quasiclassical spectra for plane waves:
\begin{eqnarray} \label{eq:tr_spinwaves_qc}
\nonumber
 c_\perp^2\ k^2 + (c_\parallel^2-c_\perp^2) ({\bf k\cdot\hat l})^2
+ \frac12 \OmegaB^2 \sin^2\beta_n
&=& \omega(\omega-\omegaL),\\
\label{eq:lo_spinwaves_qc}
 C_\perp^2 k^2 + (C_\parallel^2-C_\perp^2) ({\bf k\cdot\hat l})^2
+ \OmegaB^2 \cos^2\beta_n &=& \omega^2,
\end{eqnarray}
where $\beta_n$ is an angle between ${\bf\hat n}$ and ${\bf H}$,
the orbital anisotropy axis $\hat l_j = R_{aj} S^0_a$
and the spin wave velocities are introduced as
\begin{equation}
\begin{aligned}
&c_\perp^2 = \frac{\gamma^2\Delta^2}{\chi_B}(K-K'/2),\quad
c_\parallel^2 = \frac{\gamma^2\Delta^2}{\chi_B} K,
\\
&C_\perp^2 = \frac{\gamma^2\Delta^2}{\chi_B} K,\quad
C_\parallel^2 = \frac{\gamma^2\Delta^2}{\chi_B} (K-K').
\end{aligned}
\end{equation}

The spin wave velocities are anisotropic, they have different values if
the wave propagates in the direction of~$\bf\hat l$ or in the
perpendicular direction. The second line of Supplementary
equation~(\ref{eq:tr_spinwaves_qc}) describes a longitudinal wave, the
light Higgs mode with a ``relativistic'' spectrum
\begin{equation}
\omega_{\rm Higgs} = \sqrt{\OmegaB^2 + (C\ k)^2}.
\label{eq:omlong}
\end{equation}
The first line of Supplementary equation~(\ref{eq:tr_spinwaves_qc})
describes two modes of transverse waves, optical and acoustic magnons,
with spectra of the form:
\begin{equation}
\omega_{\rm opt} = \frac{\omegaL}{2} +
\sqrt{\left(\frac{\omegaL}{2}\right)^2 + (c\ k)^2},\quad
\omega_{\rm ac} = -\frac{\omegaL}{2} +
\sqrt{\left(\frac{\omegaL}{2}\right)^2 + (c\ k)^2}.
\label{eq:omtr}
\end{equation}
In Supplementary equations (\ref{eq:omlong}) and (\ref{eq:omtr}) the
effects of anisotropy and of the spin-orbit interaction are omitted for
simplicity.

\subsection*{Little Higgs field for spin waves in $^3$He-B}

Let us introduce a vector field
\begin{equation}\label{vector}
{\bf n}=\hat {\bf n}\sin \theta/2\,. 
\end{equation}
The spin-orbit interaction~(\ref{eq:en_d_nt}) provides a ``Mexican Hat'' potential for the ${\bf n}$-field
\begin{equation}\label{FD}
\FSO=\Lambda (|{\bf n}|^2 - n_0^2)^2\,.
\end{equation}
where $n_0^2=5/8$ and parameter
$\Lambda=\frac{32}{15}\frac{\chi_B}{\gamma^2}\Omega_\mathrm{B}^2$.

In the terminology of particle physics the ${\bf n}$-vector serves as
the ``little Higgs'' field. In the vacuum states the amplitude of the
field is fixed, $|{\bf n}|=n_0$,  while they are degenerate with respect
to the orientation of~$\hat{\bf n}$. The broken $SU(2)$ symmetry leads
to two Nambu-Goldstone modes  (propagating oscillations of the
orientation of~$\hat{\bf n}$), and one light Higgs mode (propagating
oscillations of the amplitude $|{\bf n}|$  around $n_0$).  These  three
modes comprising the little Higgs field are similar  to the bosonic
sector of Standard Model, where also the  $SU(2)$ symmetry is
instrumental. This low energy sector  of Standard Model contains the NG
modes (the gauge bosons) and one ``light Higgs'' (the 125 GeV Higgs
boson). Our two Nambu-Goldstone spin-wave modes correspond to the
doublet of the W-bosons. The spectrum of the spin wave modes in $^3$He-B
splits in magnetic field into acoustic and optical modes. The similar
splitting is discussed for the spectrum of the  W-bosons in magnetic
field (see e.g.\ Ref.~\cite{Chernodub2014}). Moreover, in strong
magnetic fields the Bose condensation of the W-bosons is expected, which
is similar to the  Bose condensation of optical magnons.


\section*{SUPPLEMENTARY NOTE 2}

\subsection*{Resonance condition for acoustic magnons}

Let's consider an excitation of acoustic magnons with frequency
$N\,\omegaL$. Here $N=1/2$ corresponds to the parametric exitation,
$N=1,2\ldots$ to the exitation of acoustic magnons with frequency
$\omegaL, 2\,\omegaL, \ldots$. The resonances observed in the experiments
correspond to standing waves in the cylindrical cell with radius~$R$.
For the short spin waves the resonances can be treated in the
quasiclassical approximation:
\begin{equation} \label{eq:radial_quant}
2\int_0^R k_r\ dr = n\pi\,,
\end{equation}
where $k_r$ is the classical trajectory along the cell diameter and $n$
is integer quantum number. This quantization corresponds to the wave modes
in cylinder with high radial and small azimuthal quantum numbers.

The effect of the spin-orbit interaction on the spectrum of short-wave acoustic
magnons can be neglected, but anisotropy of wave velocity is important.
The ratio of the velocities for~$\bf k\parallel \hat l$ and~${\bf
k}\perp \hat {\bf l}$ is $c_\parallel/c_\perp \approx \sqrt{4/3}$.
Substituting the transverse magnon spectrum~(\ref{eq:tr_spinwaves_qc})
without the spin-orbit term into~(\ref{eq:radial_quant}) and taking into
account that $\omega=N\omegaL$ we get
\begin{equation}\label{eq:excitation}
\omegaL = \frac{1}{\sqrt{N(1+N)}}\ \frac{\pi n\,c}{2R}
\end{equation}
where~$c$ is a harmonic mean velocity in the non-uniform texture:
\begin{equation}
1/c = \frac{1}{R}\int_0^R \left(c_\perp^2
 + (c_\parallel^2 - c_\perp^2)\ \sin^2\beta_l(r)
 \right)^{-1/2}dr,
\end{equation}
and $\beta_l$ is an angle between~$\hat {\bf l}$ and~$\bf H$.

The distance between the resonances is:
\begin{equation}\label{eq:simple_periods1}
\delta f_N = \frac{1}{2\pi}\ \frac{\partial \omegaL}{\partial n} =
\frac{1}{\sqrt{N(1+N)}}\ \frac{c}{4R}\,.
\end{equation}

Note that the spin wave spectra~(\ref{eq:tr_spinwaves_qc}) have been obtained
with the assumption of zero coupling between transverse and longitudinal
modes ($\omega^2/\OmegaB^2\gg 1$ or $\beta_n\ll 1$). In our experiment
this condition is approximately valid for directly excited magnons with $\omega >
2\OmegaB$. We use the same approximation also for parametrically excited magnons with
$\omega\approx\OmegaB$. This is probably the reason why the agreement of
the experimental data with Eq.~(\ref{eq:simple_periods1}) is
much better for the directly excited magnons.

\section*{SUPPLEMENTARY NOTE 3}

\subsection*{Trap for magnon quasiparticles}


In the case of optical magnons with~$\omega\approx\omegaL$, localized in
the center of the cell, where~$\bf\hat n$ is almost parallel to~$\bf H$,
equation~(\ref{eq:ham_eq3}) can be rewritten in a form of a Schr\"odinger
equation for magnon quasiparticles, where complex value
$s_+ = \frac1{\sqrt2}(S_x+iS_y)$ plays role of the wave function and precession
frequency~$\omega$ plays role of the energy. Effect of texture on the gradient terms
is neglected here because it adds only a small correction to the total gradient
energy.
\begin{equation}\label{eq:schred}
\left[
- \frac{c_\perp^2}{\omegaL}\ (\nabla_x^2+\nabla_y^2)
- \frac{c_\parallel^2}{\omegaL}\ \nabla_z^2
+ \frac{\Omega_B^2}{2\omegaL} \sin^2\beta_n + \omegaL
\right] s_{+} =
\omega\ s_{+}
\end{equation}
Non-uniform potential for magnons is formed by the order parameter
texture and the magnetic field ($\beta_n$ and~$\omegaL$ parameters).
\begin{equation}
U = \frac{\OmegaB^2}{2\gamma H} \sin^2\beta_n + \omegaL.
\end{equation}

In our setup the potential has a quadratic minimum in the center of the
sample: in the flare-out texture angle~$\beta_n$ near the sample
axis is linear, $\beta_n = \beta_n' r$ and magnetic field of the longitudinal coil
has also quadratic profile near the center.

We use pulsed NMR to populate a few lowest levels in this harmonic trap.
If the number of magnons in the system is small enough, interaction
between the levels is negligible and the excited states can be resolved
in the measurements independently. If the magnon population is above a
certain threshold, they collapse to the  ground state and form a
Bose-Einstein condensate.

From the spectra of the magnon levels in the trap we can find
values of spin-wave velocities~$c_\perp$ and~$c_\parallel$.
This work is presented in Ref.~\onlinecite{SpinWaveVel}.


\end{document}